\documentclass[aps,pre,preprint,groupedaddress]{revtex4}

\usepackage{graphicx}
\usepackage{latexsym}

\begin{document}

%\preprint{}

\title{
Folding of the triangular lattice
in a discrete three-dimensional space:
Density-matrix-renormalization-group study
}

\author{Yoshihiro Nishiyama}
%\thanks{}
\affiliation{Department of Physics, Faculty of Science,
Okayama University, Okayama 700-8530, Japan.}

\date{\today}

\begin{abstract}
Folding of the triangular lattice 
in a discrete three-dimensional space
is investigated numerically.
Such ``discrete folding'' has come under through theoretical investigation,
since Bowick and co-worker introduced it
as a simplified model for the crumpling of
the phantom polymerized membranes.
So far, it has been analyzed with the hexagon approximation
of the cluster variation method (CVM).
However,
the possible systematic error of the approximation was not fully estimated;
in fact, it has been known that
the transfer-matrix calculation is limited in the tractable strip widths $L \le 6$.
Aiming to surmount this limitation,
we utilized the density-matrix renormalization group.
Thereby, we succeeded in treating strip widths up to $L=29$
which admit
reliable extrapolations to the thermodynamic limit.
Our data indicate an onset of a discontinuous crumpling transition
with the latent heat substantially larger than the 
CVM estimate.
It is even larger than the latent heat of the planar (two dimensional) folding,
as first noticed by the preceding CVM study.
That is, contrary to our naive expectation,
the discontinuous character of the transition
is even promoted by the enlargement of the embedding-space dimensions.
We also calculated
the folding entropy, which appears to lie within
the best analytical bound obtained previously via combinatorics arguments.
\end{abstract}

%\keywords{}

\maketitle

\section{\label{section1}Introduction}

Statistical mechanics of membranes is regarded as a natural 
extension of that of polymers.
The interplay between the extended geometry and the
thermal fluctuations has provided even richer subjects,
leading to a very active area of research 
\cite{Nelson89,Nelson96,Bowick01}.
Depending on fine details of microscopic interactions,
the behaviors of membranes are distinguished
into a number of subgroups:
In the case where the constituent molecules are diffusive,
the membrane cannot support a shear, and the elasticity
is governed only by the bending rigidity \cite{Canham70,Helfrich73}.
Such a membrane is called fluid (lipid) membrane, and it
is always crumpled irrespective of temperatures.
On the contrary, provided that fixed inter-molecular
connectivity is formed via polymerization,
the in-plane strain is subjected to finite shear moduli.
Such a membrane is called polymerized (tethered) membrane.
Contrary to the fluid membrane,
the polymerized membrane is flattened macroscopically
for sufficiently large rigidities, or equivalently,
at low temperatures \cite{Nelson87}.
(Hence, the crumpling transition separates the flat and the crumpled phases.)
The flat phase is characterized by the long-range orientational
order of the surface normals.
It is rather exceptional that for such a two-dimensional
system, continuous symmetry is broken spontaneously.
To clarify this issue, a good deal of analyses
have been reported so far.
However, there still remain controversies whether
the crumpling transition belongs to a continuous transition
\cite{Kantor86,Kantor87,Baig89,Ambjorn89,Renken96,Harnish96,Baig94,%
Bowick96b,Wheather93,Wheater96,David88,Doussal92,Espriu96}
or a discontinuous one accompanying appreciable latent heat
\cite{Paczuski88,Kownacki02}.

Meanwhile, a unique alternative approach to this problem was
initiated
by Kantor and Jari\'c \cite{Kantor90},
 who formulated a discretized version of the polymerized membrane.
To be specific, they considered a triangular lattice (network),
whose joint angle $\theta$ (relative angle between adjacent plaquettes)
is taken from either $\theta=\pi$ (no fold) or
$\theta=0$ (complete fold).
Because the embedding space is two dimensional, 
this model is referred to as planar folding.
(In the theory, we consider an unstretchable sheet without
self-avoidance. 
Hence, the model is regarded as a discretized version of the 
phantom polymerized membrane with infinite strain moduli.)
Owing to this discretization,
the problem reduces to an Ising model;
the resultant spin model is, however, subjected to a constraint
which makes the problem highly nontrivial.
Nevertheless, owing to the discretization, we are able to
resort to numerous techniques developed
in the studies of the spin models.
Actually, both analytical (cluster variation) and 
numerical (transfer matrix) approaches have been utilized 
successfully \cite{DiFrancesco94b,Cirillo96b}.
Thereby, it turned out that
the crumpling transition is discontinuous,
and the discrete character of the transition is maintained
even in the presence of a perturbation such as the symmetry
breaking field (coupling to the spin variables).

One may be tempted to ask how this conclusion is
sensitive to the enlargement of the embedding space dimensions.
It is conceivable that the transition disappears
altogether
or that the order of the
transition changes.
Motivated by such an idea,
Bowick and co-workers enlarged
the embedding space to three dimensions 
\cite{Bowick95,Cirillo96,Bowick96,Bowick97}.
Correspondingly,
the joint angles are now taken
from four distinctive values; we will explain the details afterward.
However, the enlargement of the configuration space
overwhelms the computer-memory space in the diagonalization 
procedure of the transfer matrix.
The tractable strip widths $L$ are limited within the size $L=6$ which might be
rather insufficient so as to examine the possible systematic error of the
cluster-variation approximation.

In this paper, we surmount the limitation through resorting to the
density-matrix renormalization group \cite{White92,White93,Nishino95}.
This technique has proved to be successful even in the problem of the
soft materials such as the fluid membranes \cite{Nishiyama02,Nishiyama03}.
Taking the advantage that the technique allows us to treat very large
system sizes, we investigate the bulk properties of the discrete folding
in detail.
Our first-principles data indicate that the crumpling transition is
discontinuous, and the latent heat is considerably larger than the 
cluster-variation estimate.
It is even larger than the latent heat of the planar folding,
as first noted by the cluster-variation-method study \cite{Cirillo96}. 
That is, contrary to our naive expectation, the discontinuous character of the 
transition is even pronounced by the enlargement of the embedding-space
dimensions.
We will also calculate the folding entropy that has a close connection with
a combinatorics problem in mathematics.
Our result lies within the best analytical bound obtained via combinatorics
arguments \cite{Bowick95}.

The rest of this paper is organized as follows.
In the next section, we explicate the three-dimensional 
discrete folding.
The main goal is to set up an expression for the transfer-matrix element
\cite{Bowick95}.
In Sec. \ref{section3}, we present our scheme for
the diagonalization of the transfer matrix with the
density-matrix renormalization group.
A preliminary simulation result for a performance check 
is shown as well.
In Sec. \ref{section4}, we perform extensive simulations.
Our first-principles data are compared with the preceding
cluster-variation results \cite{Cirillo96}.
The last section is devoted to the summary and discussions.

\section{\label{section2}
Construction of the transfer matrix
for the three-dimensional discrete folding
\cite{Bowick95}}

In this section, we set up an expression for the transfer matrix
of the three-dimensional discrete folding \cite{Bowick95}.
The transfer matrix is treated numerically in the succeeding sections.
To begin with, we explain how the triangular lattice
(network) is embedded onto the face-centered-cubic (FCC) lattice:
The FCC lattice is viewed as a packing of the three-dimensional
space with octahedrons and tetrahedrons, whose vertices are shearing the
lattice points of the FCC lattice; see Fig. 3 of Ref. \cite{Bowick95}.
Because faces of these polygons are all equilateral triangles,
a sheet of the triangular lattice (network) can be embedded onto the
FCC lattice so as to wrap these polygons in arbitrary ways.
Because there are two types of polygons,
the relative fold angles $\theta$ between the adjacent plaquettes (triangles)
are now taken from four possibilities.
That is, in addition to the cases of ``no fold'' ($\theta=\pi$)
and ``complete cold'' ($\theta=0$) that are already incorporated in the planar
folding, we consider ``acute fold'' ($\theta=acos(1/3)$) and
``obtuse fold'' ($\theta=acos(-1/3)$).
(Note that we ignore the distortions of plaquettes (triangles).
Hence, the limit of large strain moduli is assumed {\it a priori}.)

\begin{figure}
\includegraphics{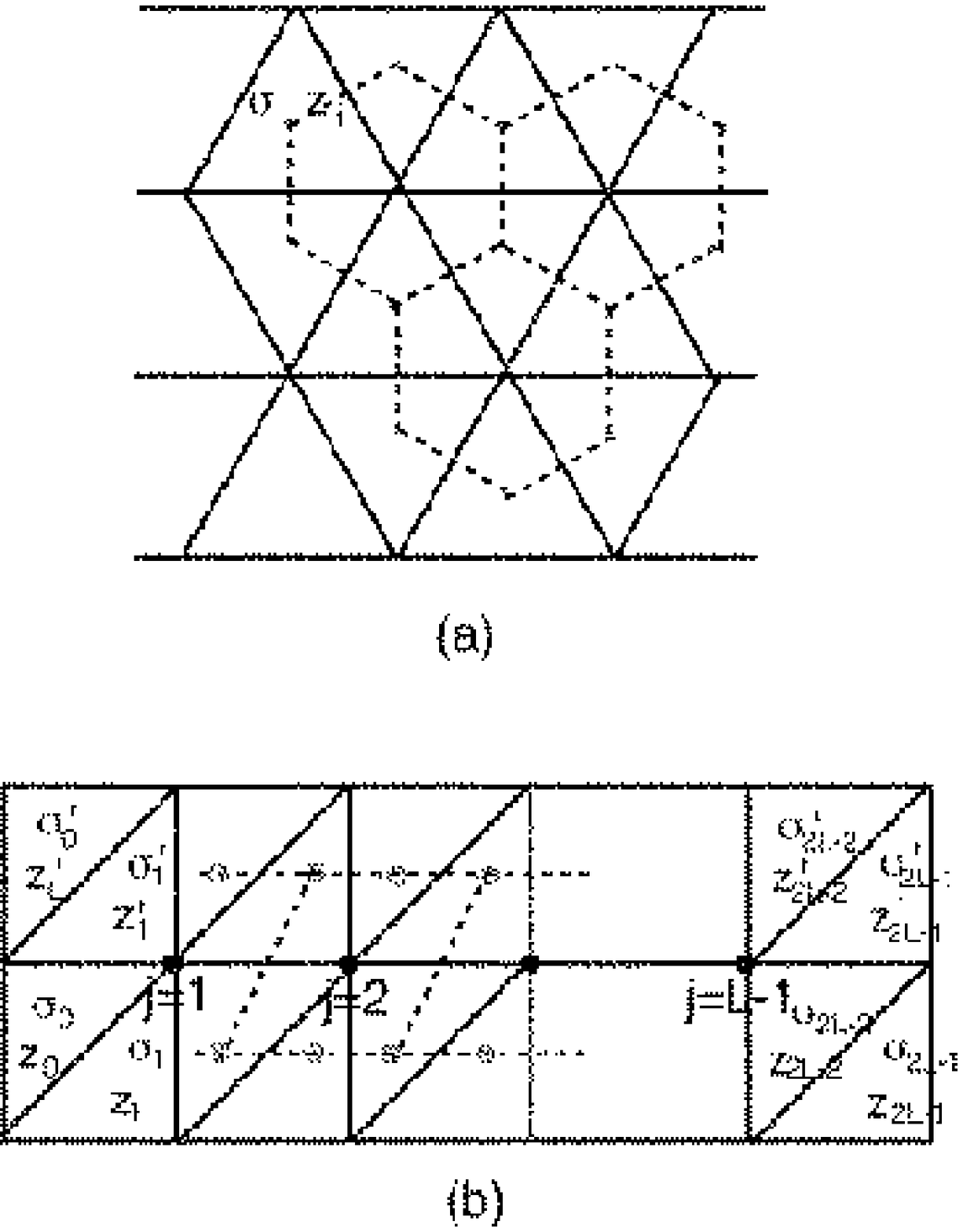}%
\caption{\label{figure1}
(a) We consider a discrete folding of the triangular lattice
embedded in the three-dimensional space.
In order to specify the fold angle, we place two types of
Ising variables such as $z_i$ and $\sigma_i$ at each triangle
rather than at each joint (gauge rule \cite{Bowick95}).
Hence, hereafter, we consider a spin model 
defined on the dual (hexagonal) lattice.
(b) A construction of the transfer matrix.
The row-to-row statistical weight yields the transfer-matrix element.
The explicit formula is given by Eq. (\ref{transfer_matrix}).
The transfer matrix is diagonalized with the density-matrix renormalization
group; see Fig. \ref{figure2}.
}
\end{figure}

The above discretization leads to an Ising-spin representation for the
membrane folding.
An efficient representation, the so-called gauge rule, 
reads as follows \cite{Bowick95}:
We place two types of Ising variables,
namely, $\sigma_i$ and $z_i$,
at each triangle (rather than each joint); see Fig. \ref{figure1} (a).
The gauge rule states that the sign change of the Ising variables
with respect to neighboring triangles specifies the joint angle:
That is, provided that the sign of the $z_i$ variables 
changes ($z_1 z_2=-1$), the joint angle is either acute or
obtuse fold.
Similarly, if $\sigma_1 \sigma_2=-1$ holds, 
the relative angle is either complete or obtuse fold.
Note that the above rules specify the joint angle unambiguously.

Let us summarize the points:
We introduced statistical variables placed at each triangle.
Therefore, the resultant spin model is defined on the hexagon lattice,
which is dual to the triangular lattice.
Hence, the transfer-matrix strip looks like that drawn in Fig. \ref{figure1} (b).
The row-to-row statistical weight $T_{\{\sigma_i,z_i\},\{\sigma_i',z_i'\}}$
yields the transfer-matrix element.
However, according to Ref. \cite{Bowick95}, the
Ising variables are not quite independent, and
constraints should apply to the spins around each hexagon.
The transfer-matrix element is given by the following form
with extra factors that enforce the constraints \cite{Bowick95};
\begin{equation}
\label{transfer_matrix}
T_{ \{z,\sigma\},\{z',\sigma'\} } = 
    ( \prod_{j=1}^{L-1} U_j V_j ) \exp(-H)      ,
\end{equation}
with,
\begin{equation}
\label{constraint1}
U_j= \delta(
 \sigma_{2j-2}+\sigma_{2j-1}+\sigma_{2j}
    +\sigma'_{2j-1}+\sigma'_{2j}+\sigma'_{2j+1}\ mod\ 3,0)   ,
\end{equation}
and,
\begin{equation}
\label{constraint2}
V_j=  \prod_{c=1}^{2} \delta(
\alpha_c(z_{2j},z_{2j-1},z_{2j-2},z'_{2j-1},z'_{2j},z'_{2j+1})\ mod\ 2,0)
                            .
\end{equation}
Here, $\delta(m,n)$ denotes Kronecker's symbol, and $\alpha_c$
is given by,
\begin{equation}
\alpha_c(z_1, \dots ,z_6)=\sum_{i=1}^6
\frac{1}{2}(1-z_i z_{i+1}) \delta(c_0+\sum_{j=1}^i \sigma_i - c\ mod\ 3,0)  .
\end{equation}
(These constraints, Eqs. (\ref{constraint1}) and (\ref{constraint2}),
allow 96
spin configurations around each hexagon \cite{Bowick95}.
In that sense,
the resultant model is regarded as a 96-vertex model.)
The Boltzmann factor $\exp(-H)$ in Eq. (\ref{transfer_matrix}) is due to
the bending-energy cost; we set the temperature $T$ as the unit of energy; namely, $T=1$.
As usual,
the bending energy is given by the inner product of the surface
normals of adjacent triangles.
Hence, the bending energy is given by the formula;
\begin{equation}
\label{elastic_energy}
H=  - 0.5  \sum_{\langle i j \rangle} 
 K \cos \theta = - 0.5 \sum_{\langle i j \rangle}
              \frac{1}{3}K \sigma_i \sigma_j (1+2 z_i z_j)  ,
\end{equation}
with the bending rigidity $K$.
Here, the summation $\sum_{\langle ij \rangle}$ runs over all possible nearest-neighbor
pairs $i$-$j$ around each hexagon.  
(The overall factor $0.5$ is intended to reconcile the double counting.)

The above completes the transfer-matrix construction
for the discrete folding.
The size of the matrix grows exponentially in the form $\propto 16^L$ with
the system size $L$; the definition of $L$ is shown in Fig. \ref{figure1} (b).
Apparently,
it is an involving task to diagonalize the matrix numerically;
so far, the width $L=6$ has been achieved with the conventional
full-diagonalization scheme \cite{Bowick95,Bowick96,Bowick97}.
Moreover, the open boundary condition must be imposed
in order to release the constraints from the boundaries;
otherwise, the constraints are too restrictive.
The boundary effect deteriorates the data so that the data
acquire severe finite-size corrections.
In the following section, we propose an alternative diagonalization
scheme which enables us to treat large system sizes.

\section{\label{section3}
Numerical simulation algorithm:
An application of the
density-matrix renormalization group to the discrete folding}

As is mentioned in the preceding section, the diagonalization 
of the transfer matrix (\ref{transfer_matrix}) is 
cumbersome because of its overwhelming matrix size.
Here, we propose an alternative approach through
resorting to the density-matrix renormalization group
\cite{White92,White93,Nishino95}.
The method allows us to treat large system sizes.
Our algorithm is standard, and we refer the reader to consult with
the text \cite{Peschel99} for technical details.
In the following, we outline the algorithm,
placing an emphasis on the changes specific to this problem.
We will also present a preliminary simulation result
in order to demonstrate the performance.

Basic idea of the density-matrix renormalization group is simple.
It is a sort of computer-aided real-space decimation procedure
with the truncation error far improved compared with the previous ones.
Sequential applications of the procedure enable us to reach
very long system sizes.
One operation of the real-space-decimation procedure is
depicted In Fig. \ref{figure2}.
Through the decimation, the block states and the adjacent spin variables
(hexagon)
are renormalized altogether into a new block states.
The number of states for a renormalized block is kept bounded
within $m$; the parameter $m$ sets the simulation precision.
Such $m$ states are chosen from the eigenstates
with significant statistical weights (eigenvalues) $\{ w_\alpha \}$ 
of a 
(local) density matrix \cite{Peschel99};
this is an essential part of the density-matrix renormalization group 
\cite{White92}, and the name comes from this.

This is a good position to address a few remarks regarding
the changes specific to the present problem.
First, in our simulation, in order to reduce the truncation error
of the real-space decimation,
we adopted the ``finite-size method'' \cite{Peschel99}.
We managed 3 sweeps at least, and confirmed that a good convergence
is achieved
in the sense that the error is negligible compared with the finite-size corrections.
Secondly, we have applied a magnetic field $H$ of the strength $H=2$
at an end of the
transfer-matrix strip.
This trick was first utilized in Ref. \cite{Bowick97},
and it aims to split off the degeneracy caused by the
trivial four-fold symmetry ($\sigma=\pm1,z=\pm1$) of the overall membrane orientation.

We turn to a demonstration of the performance of the scheme.
In Fig. \ref{figure3}, we present the distribution of the statistical weights
$\{ w_\alpha \}$ for the bending rigidity $K=0.1$,
the number of states kept for a block $m=15$, and the system size $L=14$.
We see that the statistical weight drops very rapidly.
In fact, the 15th state exhibits very tiny statistical weight
$w_{15}\approx 3 \cdot 10^{-4}$.
Hence, only significant 15 bases, for instance, are sufficient 
to attain a precision of $\sim 3 \cdot 10^{-4}$;
note that the statistical weight of the discarded
states, namely, $w_\alpha$ ($\alpha>m$), indicates an amount of the truncation error.
In our simulation, we remain, at most, $m=20$ bases for
a renormalized block, and thus, our simulation result is
``exact'' in a practical sense.
Of course, there are finite-size corrections that are severer
than the truncation error, and they are to be considered separately.

\section{\label{section4}Numerical results}

In this section, we present the numerical results.
We first survey overall features for a wide range of 
the bending rigidity $K$.
Guided by the findings, we perform large-scale simulations,
aiming to investigate
the crumpling transition and the folding entropy.

\subsection{Preliminary survey and technical remarks}

In Fig. \ref{figure4}, we plotted the internal energy $E$
and the free energy $F$
for a wide range of the bending rigidity $K$.
Both $E$ and $F$ are normalized so as to represent the 
one-unit-cell quantities.
In order to evaluate $F$ reliably, we carried out a trick,
which we explain afterward.
Technical informations on the simulation parameters
are listed in the figure caption.
From the figure, we see a clear signature of an onset of
the crumpling transition at around $K_c \sim 0.2$.
To be specific, we see
an abrupt change of the internal energy $E$, 
and the discontinuity is pronounced
as the system size $L$ is enlarged.
This tendency suggests that the singularity belongs to the
first-order transition.
Such tendency was observed in the preceding full-diagonalization
calculation study for $L \le 6$ \cite{Bowick96}.
Such a discontinuous crumpling transition was predicted by the
cluster-variation analysis as well \cite{Cirillo96}, and we pursue this issue further 
in the next subsection.
In Inset, we have plotted the entropy per unit cell $S$.
From the plot, we see that a large amount of entropy is released 
at the transition point.
This fact tells that the membrane is flattened macroscopically
in the large-$K$ regime (flat phase).
As a matter of fact, it is to be noted that the entropy vanishes 
in the flat phase; that is, the membrane undulations are
frozen up completely withstanding the thermal disturbances.
In fact, the internal energy is well fitted by the formula 
$E=-0.5 K (3 \cos 0) =-3K/2$; see the dotted line in Fig. \ref{figure4}.
This fact again tells that the surface normals of all plaquettes
are parallel to each other ($\theta=0$).
This feature is quite reminiscent of the planar folding
\cite{DiFrancesco94b}; see Introduction.
In this three-dimensional case as well, the membrane is flattened completely
in the large-$K$ phase.
In the present case, however,
the entropy release may be smeared out to some extent
by the enlargement
of the embedding space dimensions.
This issue is also explored in the next subsection.

\begin{figure}
\includegraphics{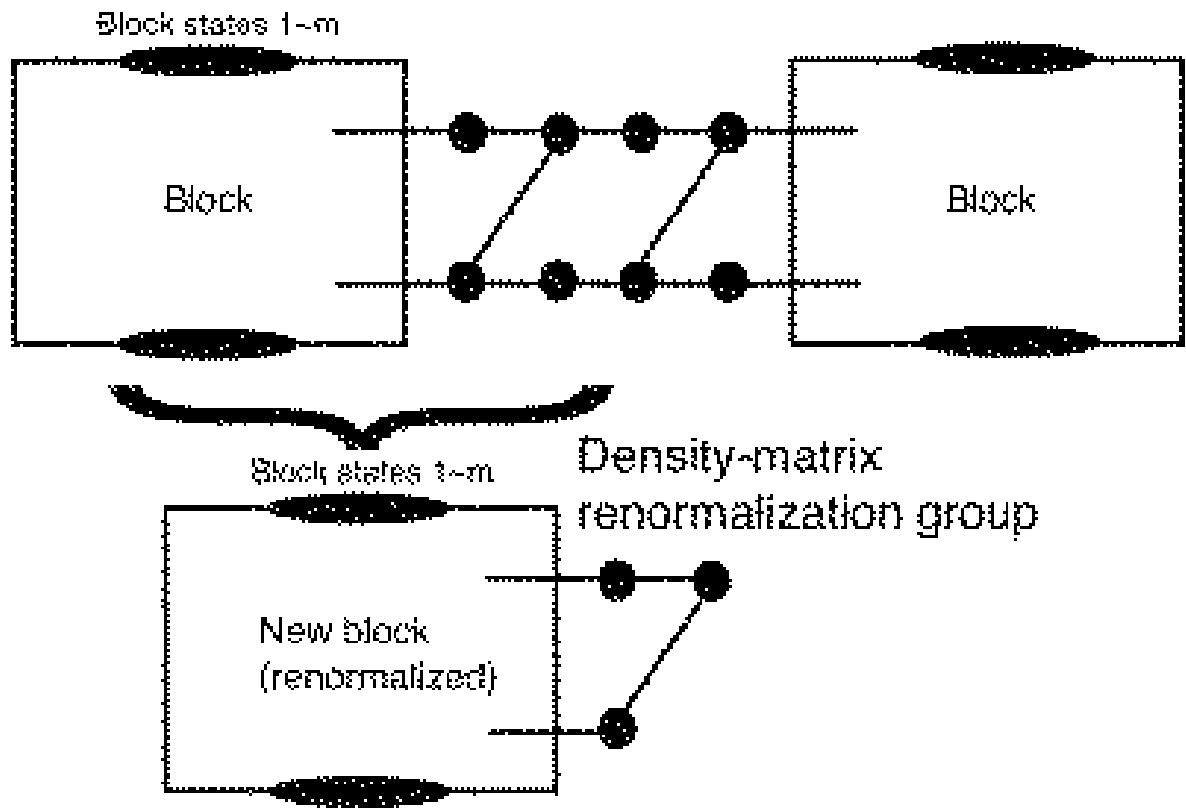}%
\caption{\label{figure2}
Schematic drawing of the density-matrix renormalization group (DMRG) 
procedure.
From the drawing, we see that through one operation of DMRG,
a ``block'' and the adjacent sites (hexagon) are 
renormalized altogether into a new renormalized ``block.''
At this time, the number of block states is retained within $m$;
see text.
In this manner,
we can diagonalize a large-scale transfer matrix
through successive applications of DMRG.
}
\end{figure}

In the following, let us mention some technical remarks:
First, we explain a trick to evaluate the free energy $F$.
Although the internal energy $E$ is calculated straightforwardly.
the calculation of $F$ requires some care.
As is well known, the total strip free energy is calculated
with the transfer-matrix diagonalization directly.
However, because of the presence of the boundaries, 
the data are deteriorated by the boundary effect.
(On the other hand, the presence of boundaries are vital 
to the numerical calculation; see Sec. \ref{section2}.)
Moreover, 
through sequential applications of the normalization,
the total 
``width'' of the strip becomes obscure.
In fact, the degrees of freedom processed in the early 
stages of normalization would be truncated out by the succeeding renormalizations.
In that sense, we should place a stress upon the midst of the transfer-matrix strip;
namely, the midst hexagon in Fig. \ref{figure2}.
The midst hexagon retains the full 
degrees of freedom (yet to be renormalized),
and the boundary effect is less influential. 
In order to extract the contribution from the midst
part, we managed a subtraction for those free energies with different strip widths
depicted in Figs. \ref{figure2} and \ref{figure5}.
The free-energy difference yields the bulk free energy containing three unit cells.
This trick is of particular importance in the succeeding subsections
where we perform numerous renormalizations.

Second, as was first noticed in Ref. \cite{Gendiar02}, 
the density-matrix-renormalization-group data exhibit a hysteresis effect.
The hysteresis is due to the fact that the
past information is encoded in the renormalized block.
In the simulation presented in Fig. \ref{figure4},
we have softened the bending rigidity $K$ gradually.
In this way, we succeeded in stabilizing the flat phase.
(Otherwise, one may not be able to realize the flat phase.
Because the flat phase is quite peculiar in the sense that it bears zero entropy,
it is very hard to reach such a state by annealing from the crumpled phase.)
In the next section, we will stabilize the crumpling phase instead,
because we are concerned in the crumpled state and the undulations
precursory of
the transition point.

\subsection{Crumpling transition}

In this subsection, we perform large-scale simulations,
aiming to clarify the nature of the crumpling transition.
In Fig. \ref{figure6}, we plotted the ``excess'' free energy 
$F+3/2K$ in the vicinity 
of the transition point.
The technical parameters are summarized in the figure cation.
As is found in the preceding subsection,
the excess free energy should vanish at the crumpling transition point,
because the the free energy in the flat phase obeys the
formula $-3K/2$.
In other words, in the plot, 
we are just comparing the free energies beside the
transition point.
Moreover, from the figure, we observe 
that the slope of the excess free energy is finite
at the transition point.
Hence, the crumpling transition is indeed discontinuous.

From the figure, we see that good convergence is achieved with 
respect to both $L$ (system size) and $m$ (number of retained bases
for a renormalized block).
In fact,
the data almost overlap each other for the parameters of 
$m \ge 15$ and $L \ge 19$.
Hence,
We estimate the transition point as $K_c=0.195(2)$.
Our simulation result shows that the previous cluster-variation estimate
$K_c=0.18548$ is remarkably precise.

Let us mention a technical remark:
As is mentioned in the above subsection, the density-matrix
renormalization group exhibits a hysteresis effect.
Hence, upon increasing the bending rigidity gradually,
we are able to retain the crumpling phase.
Taking advantage of those features rather specific to this particular
problem and the methodology,
we are able to determine the crumpling transition very precisely
in terms of the excess free energy.

As mentioned above, our data depicted in Fig. \ref{figure6}
suggest that the crumpling transition is discontinuous.
In order to estimate the amount of the latent heat,
in Fig. \ref{figure7}, we plotted the entropy $S$ in close vicinity 
of the transition point.
(The technical parameters are the same as those of Fig. \ref{figure6}.)
Note that the entropy at the transition point yields the
latent heat, because whole entropy is released at the crumpling transition.
(Again, owing to the hysteresis effect, the crumpling phase
is kept stabilized throughout the parameter range shown in the figure.)
In this way,
we estimate the latent heat as $Q=T\Delta S=0.365(5)$.

Our first-principles estimate of the latent heat may
be intriguing.
First, the present estimate appears to be substantially larger than that
of the cluster-variation study $Q_{CV,3d}=0.229(=0.18548\cdot1.237)$ \cite{Cirillo96}.
(Here, the contribution $1.237$ comes from ``jump in the energy like correlation function''
\cite{Cirillo96}.
Hence, multiplying it by the bending-rigidity modulus $K_c=0.18548$,
we obtain the latent heat; see the formula for the bending energy (\ref{elastic_energy}).)
As a matter of fact, our data of Fig. \ref{figure7}
indicate that the entropy is kept almost unchanged 
even in the vicinity of the transition point.
This fact tells that the phase transition occurs abruptly without any precursor.
(Almost all ``folding entropy'' $S(K=0)$, which is calculated in the next subsection,
is released at the transition point actually.)
We think that the constraints,
Eqs. (\ref{constraint1}) and (\ref{constraint2}),
give rise to such a notable suppression of the undulations precursory of the
crumpling transition.
In this sense, it is suspected that the constraints are not fully
appreciated by the single-hexagon-cluster approximation in the preceding work.
Second, nevertheless, it is still counterintuitive
that the latent heat is even larger than that of 
the planar folding $Q_{CV,2d}=0.11(=0.1013 \cdot 1.047)$ 
obtained by the cluster-variation method \cite{Cirillo96b,Cirillo96}.
Naively, one would expect that the latent-heat release is smeared out
by the enlargement of the embedding spatial dimensions.
Contrary to such a naive expectation,
our data indicate that the discontinuous nature
of the transition is enhanced by the enlargement of the embedding spatial 
dimensions.
The tendency observed in our simulation data
supports the claim that the crumpling transition is discontinuous
\cite{Paczuski88,Kownacki02}.

\subsection{Folding entropy}

In this subsection, we calculate the folding entropy,
namely, the entropy in the absence of the bending rigidity ($K=0$).
This quantity has a close connection with mathematics;
in the absence of the elastic term, the thermodynamics reduces to a pure
entropic problem regarding the assessment of the volume in the 
configuration space.
Hence, a combinatorics argument applies.
As a matter of fact, 
in the case of the planar folding,
the folding entropy $S_{2d}$ is obtained exactly
in the context of the coloring problem \cite{DiFrancesco94a,Baxter70,Baxter86}; namely,
$S_{2d} = \ln q_{2d}$ with
$q_{2d}=\sqrt{3}\pi \Gamma(1/3)^{3/2} /2=1.20872$.
As for the three-dimensional discrete folding, an exact bound 
$1.436 \le q_{3d} \le 1.589$ has been known
\cite{Bowick95}; 
namely, $0.361 \le S \le 0.463$.
(We see that the effective degrees of freedom per unit cell
are reduced considerably owing to the constraints 
(\ref{constraint1}) and (\ref{constraint2}).)
Both cluster variation $q=1.42805$ \cite{Cirillo96,Bowick97} and numerical estimates $q=1.43(1)$
\cite{Bowick95} lie slightly out of the analytic bound.
It may be thus desirable to make an assessment of the folding entropy
with the present first-principles simulation scheme.

In Fig. \ref{figure8}, we plotted the entropy $S$ at $K=0$
against $1/L^2$.
The simulation parameters are summarized in the figure caption. 
From the figure, we see that a good convergence is achieved with respect to
both $L$ and $m$.
Thereby, we estimate
the folding entropy in the thermodynamics
limit as $S=0.378(2)$.
Our estimate lies within the aforementioned analytic bound actually.
Remarkably enough, it is close to the lower bound (4\% error), suggesting that
the argument \cite{Bowick95} setting the lower bound is indeed capturing the essential
mechanism for the folding entropy.

\section{\label{section5}Summary and discussions}

\begin{figure}
\includegraphics{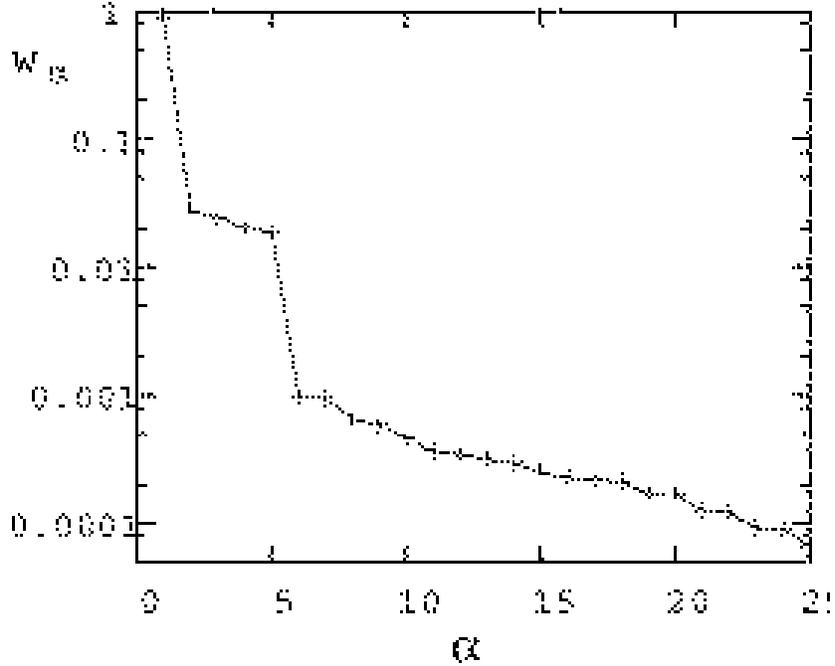}%
\caption{\label{figure3}
Distribution of the eigenvalues (statistical weights)
$\{ w_\alpha \}$ ($\alpha$, integer index) of the (local) density matrix 
\cite{White92,White93,Nishino95} is shown.
The simulation parameters are chosen from 
the bending rigidity $K=0.1$, the number of states kept for a block  $m=15$,
and the system size $L=14$.
We see that $w_\alpha$ drops very rapidly for large $\alpha$.
For instance,
the states up to only $\alpha=15$ cover the relevant (significant) bases with
appreciable statistical weights $w_\alpha > 3\cdot 10^{-4}$;
the discarded weights ($\alpha>m$)
indicate an amount of the truncation error.
In this manner, only $m$ selected (relevant) bases are retained for a
renormalized ``block'' as is shown in Fig. \ref{figure2}.
}
\end{figure}

\begin{figure}
\includegraphics{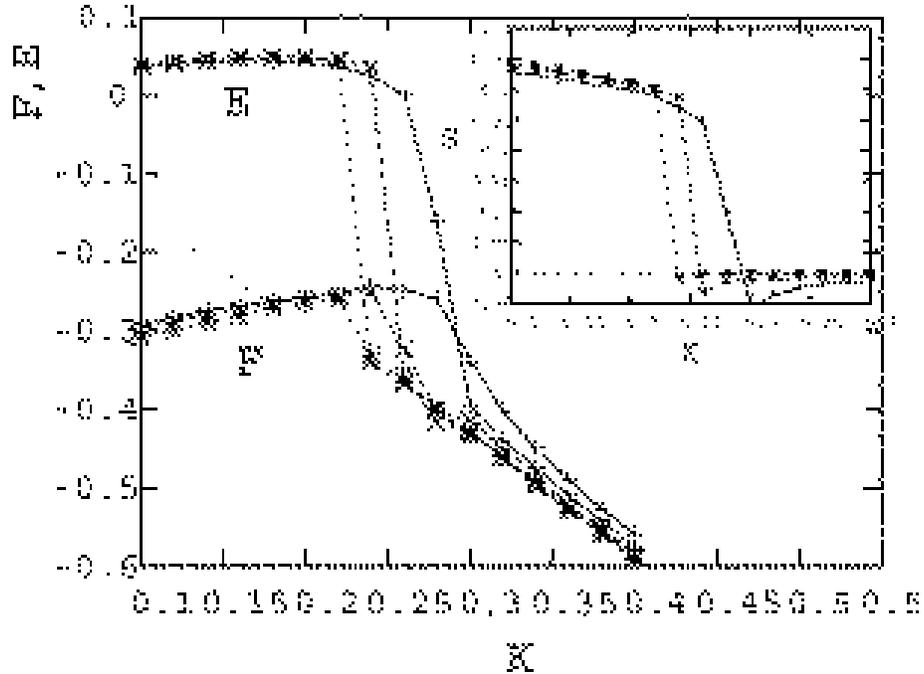}%
\caption{\label{figure4}
We plotted the free energy $F$ as well as the internal energy $E$
for the bending rigidity $K$.
The simulation parameters for each symbol are 
($+$) $L=5$ and $m=5$, ($\times$) $L=6$ and $m=5$, and 
($\ast$) $L=7$ and $m=5$.
We see an abrupt change of the internal energy $E$, 
which may indicate an onset
of the crumpling transition.
The dotted line denotes the relation $-3K/2$ which
describes the elastic energy for a completely stretched membrane.
Inset:
Entropy $S$ is plotted for the bending rigidity $K$.
In the large-$K$ phase, the entropy appears to vanish.
This result, again, indicates that the membrane is stretched
completely in the flat phase.
}
\end{figure}

\begin{figure}
\includegraphics{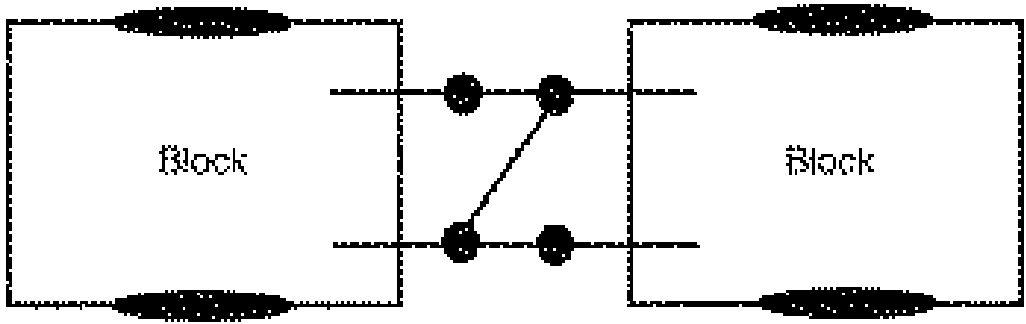}%
\caption{\label{figure5}
From the free-energy subtraction for those transfer-matrix strips
shown in the above and Fig. \ref{figure2},
we extracted the bulk contribution to the free energy
corresponding to the midst hexagon of Fig. \ref{figure2}
(containing three unit cells).
}
\end{figure}

We have investigated the three-dimensional discrete folding
by means of the density-matrix renormalization group.
The method allows us to treat large system sizes.
In fact, we succeeded in diagonalizing the transfer matrix
with the strip widths up to $L=29$.
The system sizes treated in the present study are far improved,
compared to the past limitation of $L=6$.
Taking the advantage, we could take reliable extrapolations
to the thermodynamic limit; see Fig. \ref{figure8}, for instance.
Moreover, from the figure, we confirm that the (renormalization group)
truncation error is fairly negligible $\sim 10^{-4}$.
Actually, the truncation error is far smaller than the finite-size 
corrections, and it is controlled systematically by the parameter $m$;
 see Fig. \ref{figure3} as well.

\begin{figure}
\includegraphics{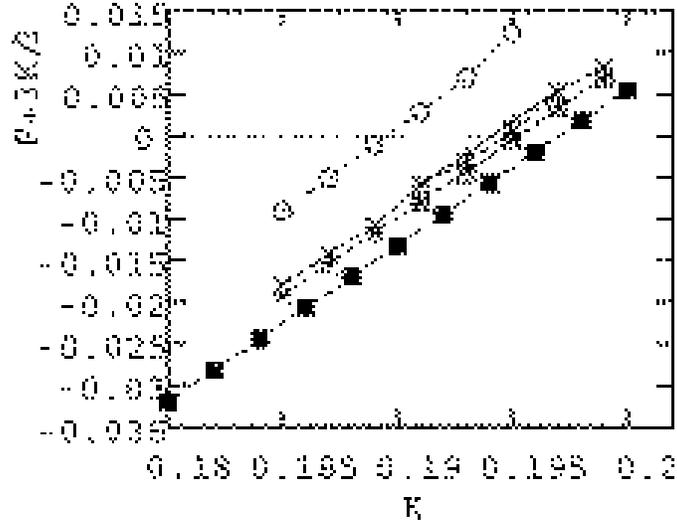}%
\caption{\label{figure6}
``Excess'' free energy $F+3K/2$ 
is plotted for the bending rigidity $K$.
The simulation parameters for each symbol are 
($+$) $L=19$ and $m=15$,
($\times$) $L=24$ and $m=20$,
($\ast$) $L=24$ and $m=15$,
($\Box$) $L=29$ and $m=15$,
(filled square) $L=24$ and $m=5$,
and
($\circ$) $L=14$ and $m=15$.
The term $-3K/2$ comes from the known exact
free energy for the flat phase.
Hence, at the crumpling transition point, 
the excess free energy should cancel out.
In this way, we obtain an estimate for the transition point
as $K_c=0.195(2)$.
(Because of the hysteresis effect \cite{Gendiar02}, the crumpled phase
is kept stabilized throughout the parameter range shown above.)
}

\end{figure}

\begin{figure}
\includegraphics{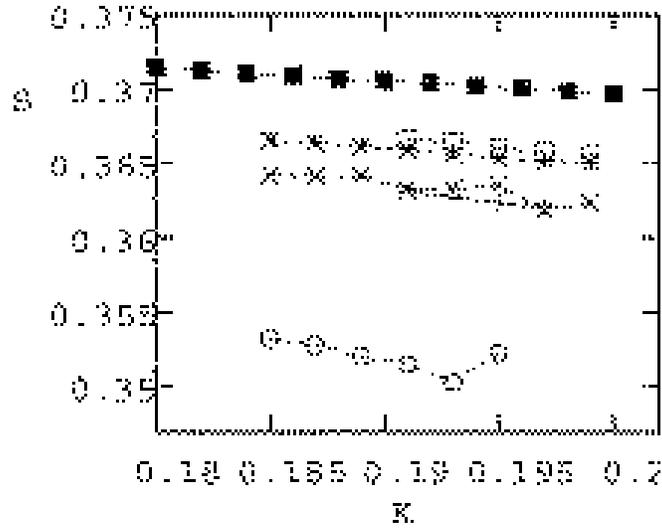}%
\caption{\label{figure7}
Entropy $S$ is plotted for the bending rigidity $K$.
The entropy right at the phase transition point $K_c=0.195(2)$  yields
the latent heat $Q=0.365(5)$; see text.
The simulation parameters are the same as those of Fig. \ref{figure6}.
(Because of the hysteresis effect \cite{Gendiar02}, the crumpled phase
is kept stabilized throughout the parameter range shown above.)}
\end{figure}

\begin{figure}
\includegraphics{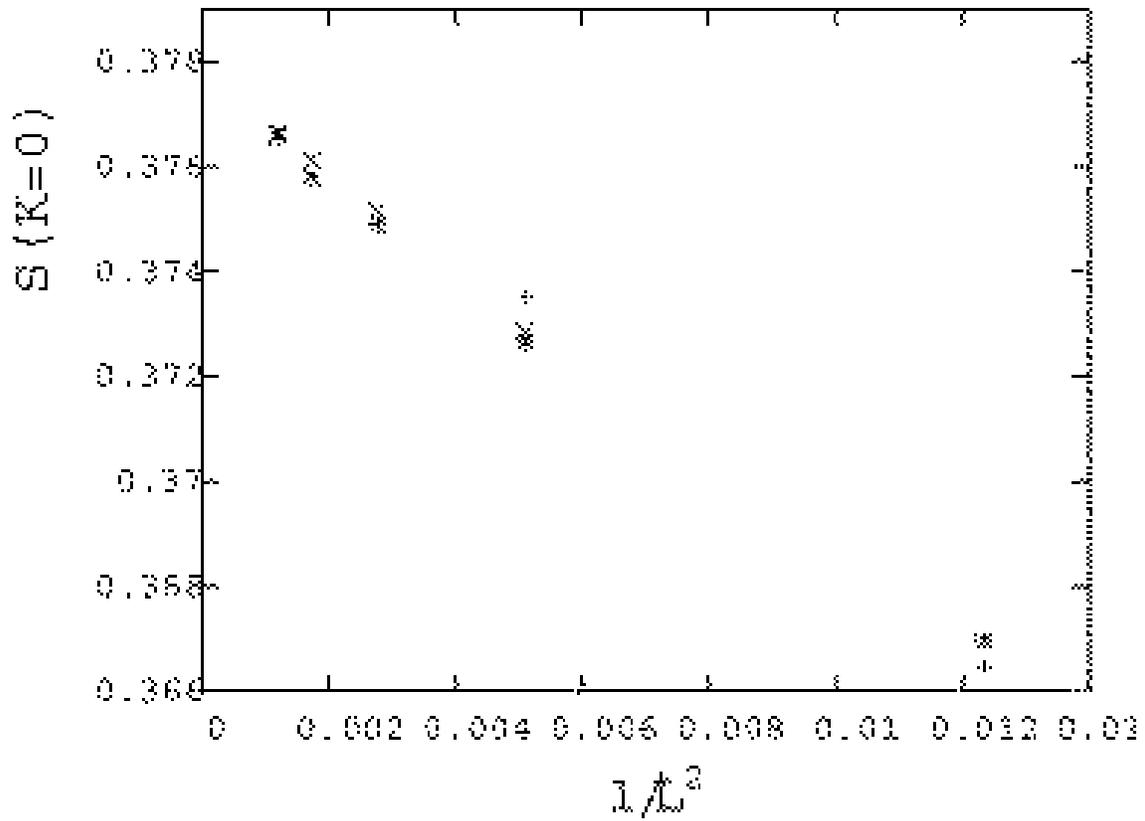}%
\caption{\label{figure8}
Folding entropy $S(K=0)$ is plotted for $1/L^2$.
The simulation parameter for each symbol is
($+$) $m=10$, ($\times$) $m=15$, and ($\ast$) $m=20$.
Taking an extrapolation to the thermodynamic limit $L\to\infty$,
we obtain an estimate $S=0.378(2)$.
Our estimate lies within the best analytic bound 
\cite{Bowick95}; see text.}
\end{figure}

Encouraged by these achievements, we turned to the analyses of the
crumpling transition and the folding entropy.
Here, we also aimed to examine the preceding analytical treatment \cite{Bowick97}
based on the cluster-variation approximation;
the approximation has been known to be reliable in the case of the
planar folding \cite{Cirillo96b}, whereas it remained unclear whether it is validated
for the three-dimensional folding.
Performing large-scale simulations,
we estimated the crumpling transition point 
$K_c=0.195(2)$ and the latent heat $Q=0.365(5)$; 
see Figs. \ref{figure6} and \ref{figure7}.
Concerning $K_c$, the numerical result is in good agreement with the cluster-variation 
estimate
$K_c=0.18548$.
As for the latent heat, however,
our simulation result appears to be 
substantially
larger than the prediction by the cluster-variation method $Q_{CV,3d}=0.229$.
Our first-principles data suggest that the constraints,
Eqs. (\ref{constraint1}) and (\ref{constraint2}),
were not fully appreciated by the (single-hexagon cluster) approximation
in the preceding study.
Namely, in reality, the undulations precursory of the
crumpling transition are suppressed considerably by the constraints.
As a matter of fact, the latent heat in the three-dimensional
folding is even larger than that of the planar folding,
as first noted by the cluster-variation study \cite{Cirillo96}.
This fact might be quite counterintuitive, because one expects that the
entropy release should be smeared out by the embedding-space enlargement.
Our first-principles simulation shows that such a naive expectation
does not hold, and in reality, the discontinuous nature of the 
crumpling transition is even promoted by the embedding-space enlargement.
(Actually, almost all folding entropy $S(K=0)$ is released abruptly at the transition point.)
In this sense, the tendency observed in our simulation data
supports the claim that the crumpling transition is discontinuous 
\cite{Paczuski88,Kownacki02}.

Let us turn to addressing 
 the entropy in the absence of the bending
rigidity $K=0$.
This quantity is often referred to as the folding entropy, and 
it has a close connection with a combinatorics problem in mathematics 
\cite{DiFrancesco94a,Baxter70,Baxter86};
note that because of $K=0$, the thermodynamics is governed solely
by the entropy (configuration-space volume).
In Ref. \cite{Bowick95}, an analytic bound $0.361 \le S \le 0.463$ was derived
via ingenious combinatorics arguments.
preceding results \cite{Cirillo96,Bowick97} are, however, slightly out of this bound.
We performed extensive simulation for the folding entropy
with the elastic constant switched off; see Fig. \ref{figure8}.
Thereby, we estimated the folding entropy as $S=0.378(2)$.
It lies within the best analytic bound, and actually,
it is almost overlapping the lower bound (4\% error).
Hence, it is suggested that the analytical argument
leading to this lower bound is capturing the essence of the folding mechanism.

Our first-principles simulation reveals that the
discontinuous character of the transition is promoted
by the enlargement of the embedding-space dimensions.
Possibly, the constraints, Eqs. (\ref{constraint1}) and (\ref{constraint2}),
give rise to such a notable discontinuity.
Hence, it would be of considerable interest to
release the constraint somehow, and see how the nature of the
transition changes.
The constraint release, however, results in a
severe increase of
the non-zero elements of the transfer matrix,
and thus, computationally demanding.
This problem is remained for the future study.

%\appendix
%\section{}

\begin{acknowledgments}
This work is supported by Grant-in-Aid for
Young Scientists
(No. 15740238) from Monbusho, Japan.
\end{acknowledgments}


\end{document}